\documentclass[preprint]{revtex4-1}
\usepackage{graphicx}

\def\be{\begin{equation}}	\def\ee#1{\label{#1}\end{equation}}
\def\ba{\begin{array}}	\def\ea{\end{array}}
\def\bea{\begin{eqnarray}}	\def\eea{\end{eqnarray}}
\def\mc{\mathcal}		\def\pa{\partial}
  \def\ci{\cite}  
\def\ra{\rightarrow}		
	
\def\N{{\mbox{\scriptsize N}}}	\def\S{{\mbox{\scriptsize S}}}
			\def\L{{\mbox{\scriptsize L}}}
\def\TOT{{\mbox{\scriptsize TOT}}} \def\S{{\mbox{\scriptsize S}}}
\def\QCD{{\mbox{\scriptsize QCD}}} 
 
\def\Re{{\rm Re}}   \def\Im{{\rm Im}}   

\def\Title#1{\begin{center} {\Large #1 } \end{center}}
\def\Author#1{\begin{center}{ \sc #1} \end{center}}
\def\Address#1{\begin{center}{ \it #1} \end{center}}

\newcommand\pubnumber{MAXLA-2/19, ``Meson''}
\newcommand\pubdate{\today}
\newcommand\pubblock{\rightline{\begin{tabular}{l} \pubnumber\\
\pubdate \end{tabular}}}
\newenvironment{Abstract}{\begin{quotation}  }{\end{quotation}}

\newenvironment{Presented}
{\begin{quotation} 
\begin{center} 
 PRESENTED AT\end{center}  
\begin{center}\begin{large}}{\end{large}
\end{center} 
\end{quotation}}

\begin{document}

\begin{titlepage}
\pubblock

\vfill
\Title{ Light and Heavy Mesons in The Complex Mass Scheme} 
\vspace{5mm}
\Author{Mikhail N. Sergeenko}
\Address{Fr. Skaryna Gomel State University, 
BY-246019, Gomel, BELARUS\\
{\rm msergeen@usa.com}}
\vfill
\begin{Abstract}
 Mesons containing light and heavy quarks are studied. 
 Interaction of quarks is described by the funnel-type potential 
with the distant dependent strong coupling, $\alpha_\S(r)$. 
 Free particle hypothesis for the bound state is developed: 
quark and antiquark move as free particles in of the bound system. 
 Relativistic two-body wave equation with position dependent particle 
masses is used to describe the flavored $Qq$ systems. 
 Solution of the equation for the system in the form of a~standing 
wave is given. 
 Interpolating complex-mass formula for two exact asymptotic eigenmass 
expressions is obtained. 
 Mass spectra for some leading-state flavored mesons are calculated. 

%
\vskip 5mm
\noindent Pacs: 11.10.St; 12.39.Pn; 12.40.Nn; 12.40.Yx\\
\noindent Keywords: 
bound state, meson, resonance, complex mass, Regge trajectory
\end{Abstract}

\vspace{3mm}
\begin{Presented}
NONLINEAR PHENOMENA IN COMPLEX SYSTEMS \\
XXVI  International  Seminar\\
Chaos, Fractals, Phase Transitions, Self-organization \\
May 21--24, 2019, Minsk, Belarus 
\end{Presented}
\vspace{5mm}
\end{titlepage}

\section{Introduction}\label{intro}
 Mesons are most numerous of hadrons in the Particle Data Group (PDG) 
tables~\ci{PDG2014}.  
 They are simplest relativistic quark-antiquark systems in case of 
equal-mass quarks, but they are not simple if quarks' masses are 
different. 
 It is believed that physics of light and heavy mesons is different; 
this is true only in asymptotic limits of large and small distances. 
 Most mesons listed in the PDG being unstable and are resonances, 
exited quark-antiquark states. 
 There are great amount and variety of experimental data and the 
different approaches used to extract the properties of 
the mesons~\ci{KlemZaits07,NussLamp02,LiMaLi04}. 

 Heavy $Q\bar Q$ mesons (quarkonia) can be considered as 
nonrelativistic (NR) bound systems and are well described by 
the~two-particle Shr\"odinger's wave equation. 
 Light $q\bar q$ states are relativistic bound states and require 
another approach. 
 Within the framework of Quantum Field Theory (QFT) the covariant 
description of relativistic bound states is the Bethe-Salpeter (BS) 
formalism~\ci{LuchSho16,BethSal08}. 
 The homogeneous BS equation governs all the bound states. 
 However, numerous attempts to apply the BS formalism to relativistic 
bound-state problems give series of difficulties. 
 Its inherent complexity usually prevents to find the exact solutions 
or results in the appearance of excitations in the relative time 
variable of the bound-state constituents (abnormal solutions), 
which are difficult to interpret in the framework of quantum 
physics~\ci{LuchSho99}. 
 Usually, calculations of hadron properties are carried out with 
the help of phenomenological and relativistic quark 
models~\ci{Morp90,EbFausGa11}. 

 Description of the heavy-light $Q\bar q$ systems in a way fully 
consistent with all requirements imposed by special relativity and 
within the framework of QFT is one of the great challenges in 
theoretical elementary particle physics~\ci{CagnAll94}. 
 For various practical reasons and applications to both QED and QCD 
some simplified equations, situated along a path of NR reduction, 
are used. 
 More valuable are methods which provide either exact or approximate 
analytic solutions for various forms of differential equations. 
 They may be remedied in three-dimensional reductions of the BS 
equation. 
 In most cases the analytic solution can be found if original 
equation is reduced to the Schr\"odinger-type wave equation. 
 The most well-known of the resulting bound-state equations is 
the one proposed by Salpeter~\ci{Salpet52}. 
 There exist many other approaches to bound-state problem. 
 One of the promising among them is the Regge method in hadron 
physics~\ci{Collin77}. 

 All hadrons and their resonances in this approach are associated 
with Regge poles which move in the complex angular momentum $J$~plane. 
 Moving poles are described by the Regge trajectories, $\alpha(s)$, 
which are the functions of the invariant squared mass $s=W^2$ 
(Mandelshtam's variable), where $W=E^*$ is the c.\,m. rest energy 
(invariant mass of two-particle system). 
 Hadrons and resonances populate their Regge trajectories which 
contain all the dynamics of hadron interaction in bound state and 
scattering regions. 

 Light and heavy mesons have been studied in a soft-wall holographic 
approach AdS/CFT~\ci{Lyubo10} using the correspondence of string 
theory in Anti-de Sitter space and conformal field theory in 
physical space-time. 
 It is analogous to the Schr\"odinger theory for atomic physics 
and  provides a precise mapping of the string modes $\Phi(z)$ in 
the AdS fifth dimension $z$ to the hadron light-front wave functions 
in physical space-time. 

 In this work we study $Q\bar q$ mesons and their excitations 
(resonances) as relativistic two-body (R2B) systems from unified 
point of view in the framework of the relativistic quantum mechanics 
(RQM)~\ci{GreinerRQM00,Dirac49}. 
 The issue is connected with two fundamental problems: 
1)~two-particle relativistic wave equation of motion and 
2)~absence of a~strict definition of the potential in relativistic 
theory. 
 Using relativistic kinematics and the correspondence principle, 
we deduce a two-particle wave equation. 
 The interaction of particles (quarks) is described by the modified 
funnel-type Lorentz-scalar Cornell potential. 
 We obtain two exact asymptotic solutions of the equation which 
are used to write the complex-mass formula for the bound system.

\section{ Relativistic two-body problem}\label{Rel2B}
 Quarkonia as quark-antiquark bound states are simplest among mesons. 
 The quarkonium universal mass formula and ``saturating'' Regge 
trajectories were derived in~\ci{MyZPhC94} and in~\ci{MyEPJC12,MyEPL10} applied for gluonia (glueballs). 
 The mass formula was obtained by interpolating between NR heavy 
$Q\bar Q$ quark system and ultra-relativistic limiting case of light 
$q\bar q$ mesons for the Cornell potential~\ci{LattBali01,EichGMR08}, 
\be 
V(r)=V_\S(r)+V_\L(r)\equiv-\frac 43\frac{\alpha_\S}r +\sigma r.
\ee{CornPot} 
 The short-range Coulomb-type term $V_\S(r)$, originating from one-gluon exchange, dominates for heavy mesons and the linear one $V_\L(r)$, which 
models the string tension, dominates for light mesons. 
 Parameters $\alpha_\S$ and $\sigma$ are directly related to basic 
physical quantities of mesons. 

 Operators in ordinary quantum mechanics (QM) are Hermitian and 
the corresponding eigenvalues are real. 
 It is possible to extend the QM Hamiltonian into the complex domain 
while still retaining the fundamental properties of a~quantum theory. 
 One of such approaches is complex quantum mechanics~\ci{BendBH}. 
 The complex-scaled method is the extension of theorems and principles 
proved in QM for Hermitian operators to non-Hermitian operators. 

 The Cornell potential (\ref{CornPot}) is a~special in hadron physics 
and results in the complex energy and mass eigenvalues. 
 Separate consideration of two asymptotic components $V_\S(r)$ and 
$V_\L(r)$ of the potential (\ref{CornPot}) for quarkonia results in 
the complex-mass expression for resonances, which in the 
center-of-momentum (c.m.) frame is ($\hbar=c=1$)~\ci{MyAHEP13,MyNPCS14}: 
\be 
\mc{M}_\N^2 = 4\left[\left(\sqrt{2\sigma\tilde{N}}
+\frac{i\tilde\alpha m}N \right)^2
+\left(m-i\sqrt{2\tilde\alpha\sigma}\right)^2\right],
\ee{CompE2n}
where $\tilde\alpha=\frac 43\alpha_\S$, $\tilde{N}=N+(k+\frac12)$, 
$N=k+l+1$, $k$ is radial and $l$ is orbital quantum numbers; it has 
the form of the squared energy 
$\mc{M}_\N^2=4\left[(\pi_\N)^2+\mu^2\right]$ of two free relativistic 
particles with the quarks' complex momenta $\pi_\N$ and masses $\mu$. 
 This formula allows to calculate in a~unified way the centered 
masses and total widths of heavy and light quarkonia.  
 In our method the energy, momentum and quark masses are {\it complex}. 

 A more complicate case are flavored $Q\bar q$ mesons. 
 A simplest example of heavy-light two-body system is the hydrogen 
($H$) atom, comprising only a~proton and an~electron which are stable 
particles. 
 This simplicity means its properties can be calculated theoretically 
with impressive accuracy~\ci{MyRelHx19}. 
 The~spherically symmetric Coulomb potential, with interaction strength 
parametrized by dimensionless coupling (``fine structure'') constant 
$\alpha$, is of particular importance in many realms of physics.  
 The $H$ atom can be used as a tool for testing any relativistic 
two-body theory, because latest measurements for transition 
frequencies have been determined with a highest 
precision~\ci{MohrTayl}.

\vspace{3mm}
 {\bf 1. The interaction potential.} The NR QM shows very good results 
in describing bound states; this is partly because the potential is 
NR concept. 
 In relativistic mechanics one faces with different kind of speculations 
around the potential, because of absence of a strict definition of 
the potential in this theory. 
 In NR formulation, the $H$ atom is described by the Schr\"odinger 
equation and is usually considered as an electron moving in 
the external field generated by the proton static electric field 
given by the Coulomb potential.  
 In relativistic case, the binding energy of an electron in a~static 
Coulomb field (the external electric field of a point nucleus of 
charge $Ze$ with infinite mass) is determined predominantly by 
the Dirac eigenvalue~\ci{MohrTayl}. 
 The spectroscopic data are usually analyzed with the use of 
the Sommerfeld's fine-structure formula~\ci{Bohm79}, 

 One should note that, in these calculations the $S$ states start 
to be destroyed above $Z=137$, and that the $P$ states being 
destroyed above $Z=274$. 
 Similar situation we observe from the result of the Klein-Gordon 
wave equation, which predicts $S$ states being destroyed above $Z=68$ 
and $P$ states destroyed above $Z=82$. 
 Besides, the radial $S$-wave function $R(r)$ diverges as $r\ra 0$. 
 These problems are general for all Lorentz-vector potentials which 
have been used in these calculations~\ci{Huang01,Bhadur95}. 
 In general, there are two different relativistic versions: 
the potential is considered either as the zero component of 
a~four-vector, a~Lorentz-scalar or their mixture~\ci{SahuAll89}; 
its nature is a~serious problem of relativistic potential 
models~\ci{Sucher95}. 

 This problem is very important in hadron physics where, for 
the vector-like confining potential, there are no normalizable 
solutions~\ci{Sucher95,SemayCeu93}. 
 There are normalizable solutions for scalar-like potentials, but not 
for vector-like. 
 This issue was investigated in~\ci{MyZPhC94,Huang01};  
it was shown that the effective interaction has to be Lorentz-scalar 
in order to confine quarks and gluons. 
 The relativistic correction for the case of the Lorentz-vector 
potential is different from that for the case of the Lorentz-scalar 
potential~\ci{MyMPLA97}. 

 The Cornell potential (\ref{CornPot}) is fixed by the two free 
parameters, $\alpha_\S$ and $\sigma$. 
 However, the strong coupling $\alpha_\S$ in QCD is a~function 
$\alpha_\S(Q^2)$ of virtuality $Q^2$ or $\alpha_\S(r)$ in configuration 
space. 
 The potential can be modified by introducing the 
$\alpha_\S(r)$-dependence, which is unknown. 
 A~possible modification of $\alpha_\S(r)$ was introduced in~\ci{MyEPJC12}, 
\be
V_\QCD(r) = -\frac 43\frac{\alpha_\S(r)}r +\sigma r,\quad 
\alpha_\S(r)=\frac 1{b_0\ln[1/(\Lambda r)^2+(2\mu_g/\Lambda)^2]},
\ee{VmodCor}
where $b_0=(33-2n_f)/12\pi$, $n_f$ is number of flavors, 
$\mu_g=\mu(Q^2)$ --- gluon mass at $Q^2=0$, $\Lambda$ is the QCD scale 
parameter. 
the running coupling $\alpha_\S(r)$ in (\ref{VmodCor}) is frozen at  
$r\ra\infty$, $\alpha_\infty=\frac 12[b_0\ln(2\mu_g/\Lambda)]^{-1}$, and 
is in agreement with the asymptotic freedom properties, i.\,e., 
$\alpha_\S(r\ra 0)\ra 0$. 

 {\bf 2. The two-body problem in RQM.} Standard relativistic approaches 
for R2B systems run into serious difficulties in solving known 
relativistic wave equations. 
 The formulation of RQM differs from NR QM by the replacement of 
invariance under Galilean transformations with invariance under 
Poincar\`e transformations. 
 The RQM is also known in the literature as relativistic Hamiltonian 
dynamics or Poincar\`e-invariant QM with direct interaction~\ci{Dirac49}. 
 There are three equivalent forms in the RQM called ``instant'', 
``point'', and ``light-front'' forms. 

 The dynamics of many-particle system in the RQM is specified 
by expressing ten generators of the Poincar\`e group, 
$\hat M_{\mu\nu}$ and $\hat W_\mu$, in terms of dynamical variables. 
 In the constructing generators for interacting systems it is customary 
to start with the generators of the corresponding non-interacting 
system; the interaction is added in the way that is consistent with 
Poincare algebra. 
 In the relativistic case it is necessary to add an interaction $V$ 
to more than one generator in order to satisfy the commutation 
relations of the Poincar\'e algebra. 

 The interaction of a~relativistic particle with the four-momentum 
$p_\mu$ moving in the external field $A_\mu(x)$ is introduced in 
QED according to the gauge invariance principle, 
$p_\mu\ra P_\mu=p_\mu-eA_\mu$. 
 The description in the ``point'' form of RQM implies that the mass 
operators $\hat M^{\mu\nu}$ are the same as for non-interacting 
particles, i.\,e., $\hat M^{\mu\nu}=M^{\mu\nu}$, and these 
interaction terms can be~presented only in the form of 
the four-momentum operators~$\hat W^\mu$~\ci{MyRQMAnd99}. 

 Consider the R2B problem in classic relativistic theory. 
 Two particles with four-momenta $p_1^\mu$, $p_2^\mu$ and 
the interaction field $W^\mu(q_1,\,q_2)$ together compose a~closed conservative system, which can be characterized by the 4-vector 
$\mc{P}^\mu$, 
\be 
\mc{P}^\mu = p_1^\mu + p_2^\mu + W^\mu(q_1,\,q_2),
\ee{Main4vec}
where the space-time coordinates $q_1^\mu$, $q_2^\mu$ and 
four-momenta $p_1^\mu$, $p_2^\mu$ are conjugate variables, 
$\mc{P}_\mu \mc{P}^\mu=\mathsf{M}^2$; here $\mathsf{M}$ is 
the system's invariant mass. 
 Underline, that no external field and each particle 
of the system can be considered as moving source of the interaction 
field; the interacting particles and the potential are a~unified 
system. 
 There are the following consequences of (\ref{Main4vec}) and 
they are key in our approach. 

 The four-vector (\ref{Main4vec}) describes {\it free motion} of 
the bound system and can be presented as, 
\bea
E = \sqrt{\mathbf{p}_1^2+m_1^2}+\sqrt{\mathbf{p}_2^2+m_2^2}
+W_0(q_1,\,q_2)=\rm{const}, \quad \label{TwoEnr}\\ 
\mathbf{P}=\mathbf{p}_1+\mathbf{p}_2
+\mathbf{W}(q_1,\,q_2)=\rm{const}, \quad \label{TwoMom}
\eea
describing the energy and momentum conservation laws. 
 The energy (\ref{TwoEnr}) and total momentum (\ref{TwoMom}) 
of the system are the constants of motion. 
 By definition, for conservative systems, the integrals (\ref{TwoEnr}) 
and (\ref{TwoMom}) can not depend on time explicitly. 
 This means the interaction $W(q_1,\,q_2)$ should not depend on 
time, i.\,e., $W(q_1,\,q_2)=>V(\mathbf{r}_1,\,\mathbf{r}_2)$. 

 It is well known that the potential as a~function in 3D-space 
is defined by the pro\-pa\-ga\-tor $D(\mathbf{q}^{\,2})$ (Green 
function) of the virtual particle as a carrier of interaction, where 
$\mathbf{q}=\mathbf{p}_1-\mathbf{p}_2$ is the transfered momentum. 
 In case of the Coulomb potential the propagator is 
$D(\mathbf{q}^{\,2})=-1/\mathbf{q}^{\,2}$; the Fourier transform 
of $4\pi\alpha D(\mathbf{q}^{\,2}$) gives the Coulomb potential, 
$V(r)=-\alpha/r$. 
 The relative momentum $\mathbf{q}$ is conjugate to the relative vector 
$\mathbf{r}=\mathbf{r}_1-\mathbf{r}_2$, therefore, one can accept 
that $V(\mathbf{r}_1,\,\mathbf{r}_2)=V(\mathbf{r})$~\ci{LuchSho99}. 
 If the potential is spherically symmetric, one can write  
$V(\mathbf{r})=>V(r)$, where $r=|\mathbf{r}|$.  
 Thus, the system's relative time $\tau=t_1-t_2=0$ 
(instantaneous interaction). 

 Equations (\ref{TwoEnr}) and (\ref{TwoMom}) in the c.m. frame are 
\bea 
\mathsf{M} = 
\sqrt{\mathbf{p}^2+m_1^2}+\sqrt{\mathbf{p}^2 +m_2^2}+\mathsf{V}(r),
\label{ClasE2B} \\
\mathbf{P}=\mathbf{p}_1 +\mathbf{p}_2 +
\mathbf{W}(\mathbf{r}_1,\,\mathbf{r}_2)=\mathbf{0}, \label{ClasM2B}
\eea
where $\mathbf{p}=\mathbf{p}_1 =-\mathbf{p}_2$ that follows from the 
equality $\mathbf{p}_1+\mathbf{p}_2=0$; this means that 
$\mathbf{W}(\mathbf{r}_1,\,\mathbf{r}_2)=0$. 
 The system's mass (\ref{ClasE2B}) in the c.m. frame is Lorentz-scalar. 
 In case of free particles ($\mathsf{V}=0$) the invariant mass 
$\mathsf{M}=\sqrt{\mathbf{p}^2+m_1^2}+\sqrt{\mathbf{p}^2+m_2^2}$ can be 
transformed for $\mathbf{p}^2$ as 
\be
\mathbf{p}^2
=\frac 1{4s}(s-m_-^2)(s-m_+^2)\equiv\mathsf{k}^2,
\ee{InvMom1}
which is relativistic invariant, $s=\mathsf{M}^2$ is the Mandelstam's 
invariant, $m_-=m_1-m_2$, $m_+=m_1+m_2$. 

 Equation (\ref{TwoEnr}) is the zeroth component of the four-vector 
(\ref{Main4vec}) and the potential $\mathsf{W_0}$ is Lorentz-vector. 
 But, in the c.m. frame the mass (\ref{ClasE2B}) is Lorentz-scalar; 
and what about the potential $\mathsf{V}$? 
 Is it still Lorentz-vector? 
 To show that the potential is Lorentz-scalar, let us reconsider 
(\ref{ClasE2B}) as follows. 
 The relativistic total energy $\epsilon_i(\mathbf{p})$ ($i=1,\,2$) 
of particles in (\ref{ClasE2B}) given by 
$\epsilon_i^2(\mathbf{p})=\mathbf{p}^2+m_i^2$ can be represented as 
sum of the kinetic energy $\tau_i(\mathbf{p})$ and the particle rest 
mass $m_i$, i.\,e., $\epsilon_i(\mathbf{p})=\tau_i(\mathbf{p})+m_i$. 
 Then the system's total energy (invariant mass) (\ref{ClasE2B}) can be 
written in the form $\mathsf{M}=\sqrt{\mathbf{p}^2+\mathsf{m}_1^2(r)}
+\sqrt{\mathbf{p}^2+\mathsf{m}_2^2(r)}$, where 
$\mathsf{m}_{1,2}(r)=m_{1,2}+\frac 12\mathsf{V}(r)$ are 
the distance-dependent particle masses~\ci{MyNDA17} and (\ref{InvMom1}) 
with the use of $\mathsf{m}_1(r)$ and $\mathsf{m}_2(r)$ takes the form, 
\be
\mathbf{p}^2 = 
K(s)\left[s-(m_+ +\mathsf{V})^2\right]\equiv\mathsf{k}^2-U(s,\,r),
\ee{InvMom2}
where $K(s)=(s-m_-^2)/4s$, $\mathsf{k}^2$ is squared invariant 
momentum given by (\ref{InvMom1}) and 
$U(s,\,r) = K(s)\left[2m_+\mathsf{V} +\mathsf{V}^2\right]$ 
is the potential function. 
 The equation (\ref{InvMom2}) is the relativistic analogy of the NR 
expression $\mathbf{p}^2=2\mu[E-V(r)]\equiv\mathsf{k}^2-U(E,r)$. 

 The equality (\ref{InvMom2}) with the help of the fundamental 
correspondence principle gives the two-particle spinless wave 
equation,
\be 
\left[\vec\nabla^2 +\mathsf{k}^2-U(s,\,r)\right]\psi(\mathbf{r})=0. 
\ee{Rel2Eq}
 The equation (\ref{Rel2Eq}) can not be solved by known methods for 
the potential (\ref{VmodCor}). 
 Here we use the quasiclassical (QC) method and solve another wave 
equation~\ci{MyMPLA97,MyPRA96}.

\section{ Solution of the QC wave equation}\label{SolQCEq}
 Solution of the Shr\"odinger-type's wave equation (\ref{Rel2Eq}) 
can be found by the QC method developed in~\ci{MyPRA96}. 
 In our method one solves the QC wave equation derivation of which 
is reduced to replacement of the operator $\vec{\nabla}^2$ in 
(\ref{Rel2Eq}) by the canonical operator $\Delta^c$ without the first 
derivatives, acting onto the state function 
$\Psi(\vec r)=\sqrt{{\rm det}\,g_{ij}}\psi(\vec r)$, where $g_{ij}$ 
is the metric tensor. 
 Thus, instead of (\ref{Rel2Eq}) one solves the QC equation, for 
the potential~(\ref{VmodCor}), 
\be 
\Biggl\{\frac{\pa^2}{\pa r^2}+\frac 1{r^2}\frac{\pa^2}{\pa\theta^2}
+\frac 1{r^2\sin^2\theta}\frac{\pa^2}{\pa\varphi^2}
+\frac{s-m_-^2}{4s}\biggl[s-\left(m_+ -\frac 43\frac{\alpha_\S(r)}r
+\sigma r\right)^2\biggr]\Biggr\}\Psi(\mathbf{r})=0.
\ee{Rel2Equa}
 This equation is separated. 
 Solution of the angular equation was obtained in~\ci{MyPRA96} by 
the QC method in the complex plane, that gives 
$\textrm{M}_l=(l+\frac 12)\hbar$, for the angular momentum eigenvalues. 
 These angular eigenmomenta are universal for all spherically symmetric 
potentials in relativistic and NR cases. 

 The radial problem has four turning points and cannot be solved by 
standard methods. 
 We consider the problem separately by the QC method for the short-range 
Coulomb term (heavy mesons) and the long-range linear term (light mesons). 
 The QC method reproduces the exact energy eigenvalues for all known 
solvable problems in quantum mechanics~\ci{MyMPLA97,MyPRA96}. 
 The radial QC wave equation of (\ref{Rel2Equa}) for the Coulomb term 
has two turning points and the phase-space integral is found in 
the complex plane with the use of the residue theory and method of 
stereographic projection~\ci{MyPRA96,MyAHEP13} that gives 
\be
\mc{M}_\N^2|_C = 
\left(m_1+m_2\right)^2\left(1-v_\N^2\right)\pm 2im_+m_-v_\N
\equiv \Re\{\mc{M}_\N^2|_C\}\pm i\Im\{\mc{M}_\N^2|_C\},
\ee{W2Coul}
where 
$v_\N=\frac 23\alpha_\infty/N$, $N=k+l+1$. 

 Large distances in hadron physics are related to the problem of 
confinement. 
 The radial problem of (\ref{Rel2Equa}) for the linear term has four 
turning points, i.\,e., two cuts between these points. 
 The phase-space integral in this case is found by the same method 
of stereographic projection as above that results in the cubic 
equation~\ci{MyNDA17}: $s^3 + a_1s^2 + a_2s + a_3 = 0$, where 
$a_1=16\tilde\alpha_\infty\sigma-m_-^2$, 
$a_2=64\sigma^2\left(\tilde\alpha_\infty^2-\tilde N^2
-\tilde\alpha_\infty m_-^2/4\sigma\right)$, 
$a_3=-(8\tilde\alpha_\infty\sigma m_-)^2$, $\tilde N=N+k+\frac 12$, 
$\tilde\alpha_\infty=\frac 43\alpha_\infty$, 
$\alpha_\infty=\alpha_\S(r\ra\infty)$. 
 The first root $s_1(N)$ of this equation gives the physical solution 
(complex eigenmasses), $\mathsf{M}_1^2|_L=s_1(N)$, for the squared 
invariant mass. 

 Two exact asymptotic solutions obtained such a way are used to derive 
the interpolating mass formula. 
 The~interpolation procedure for these two solutions~\ci{MyZPhC94} 
is used to derive the meson's complex-mass formula, 
\be
\mc{M}_\N^2 = 
\left(m_1+m_2\right)^2\left(1-v_\N^2\right)\pm 2im_+m_-v_\N 
+\mathsf{M}_1^2|_L.
\ee{W2int}
 The real part of the square root of (\ref{W2int}) defines the~centered 
masses and its imaginary part defines the~total widths, 
$\Gamma_\N^\TOT=-2\,\Im\{\mathsf{M}_\N\}$, of mesons and 
resonances~\ci{MyAHEP13,MyNPCS14}. 

 In the QC method not only the total energy, but also momentum of 
a~particle-wave in bound state is the {\em constant of motion}. 
 Solution of the QC wave equation in the whole region is written 
in elementary functions as~\ci{MyPRA96}
\be
\mathsf{R}(r) = C_n\left\{\ba{lc} \label{osol}
\frac 1{\sqrt 2}e^{|\mathsf{k}_n|r -\phi_1}, & r<r_1,\\
\cos(|\mathsf{k}_n|r -\phi_1 -\frac\pi 4), & r_1\le r\le r_2,\\
\frac{(-1)^n}{\sqrt 2}e^{-|\mathsf{k}_n|r +\phi_2}, & r>r_2,
\ea\right.\ee
where $C_n=\sqrt{2|p_n|/[\pi(n+\frac 12)+1]}$ is the normalization
coefficient, $\mathsf{k}_n$ is the corresponding eigenmomentum found 
from solution of (\ref{Rel2Eq}), $\phi_1=-\pi(n+\frac 12)/2$ and 
$\phi_2=\pi(n+\frac 12)/2$ are the values of the phase-space integral 
at the turning points $r_1$ and $r_2$, respectively. 

 The free fit to the data~\ci{PDG2014} shows a~good agreement for 
the light and heavy $Q\bar q$ meson and their resonances. 
 To demonstrate efficiency of the model we calculate the leading-state 
masses of the $\rho$ and $B^*$ meson resonances (see tables, where 
masses are in MeV). 
 Note, that the gluon mass, $m_g=416$\,MeV, and the string tension 
$\sigma=140$\,MeV$^2$ in the independent fitting are the same. 
\begin{table}[ht]
\begin{center}
\caption{The masses of the $\rho^\pm$-mesons and resonances}
\label{rho_mes}
\begin{tabular}{lllll}
\hline\noalign{\smallskip}
\ \ Meson &~~~$J^{PC}$ &~~~$\ \ E_n^{ex}$ &~~~$\ \ E_n^{th}$&
~~~Parameters in (\ref{W2int})\\
\noalign{\smallskip}\hline\hline\noalign{\smallskip}
\ \ \ $\rho\ (1S)$&~~~$1^{--}$&~~~$\ \ 776$&~~~$\ \ 776$&
~~~~~$\Lambda=488$ MeV\\
\ \ \ $a_2(1P)$&~~~$2^{++}$&~~~$\ 1318$&~~~$\ 1315$&
~~~~~$\mu_g=416$\,MeV\\ 
\ \ \ $\rho_3(1D)$&~~~$3^{--}$&~~~$\ 1689$&~~~$\ 1689$&
~~~~~$\sigma=140$\,MeV$^2$\\ 
\ \ \ $a_4(1F)$&~~~$4^{++}$&~~~$\ 1996$&~~~$\ 1993$&
~~~~~$m_d=119$\,MeV\\ 
\ \ \ $\rho\ (1G)$&~~~$5^{--}$&~~~$ ~ $&~~~$\ 2257$&
~~~~~$m_u=69$\,MeV\\
\ \ \ $\rho\ (2S)$&~~~$1^{--}$&~~~$\ 1720$&~~~$\ 1688$&~\\
\ \ \ $\rho\ (2P)$&~~~$2^{++}$&~~~$ ~ $&~~~$\  1993$&~\\
\ \ \ $\rho\ (2D)$&~~~$3^{--}$&~~~$ ~ $&~~~$\ 2257$&~\\
\noalign{\smallskip}\hline
\end{tabular}
\caption{The masses of the $B^{*0}$-mesons and resonances}
\label{Dp_mes}
\begin{tabular}{lllll}
\hline\noalign{\smallskip}
\ \ Meson &~~~$J^{PC}$ &~~~$\ \ E_n^{ex}$ &~~~$\ \ E_n^{th}$&
~~~Parameters in (\ref{W2int})\\
\noalign{\smallskip}\hline\hline\noalign{\smallskip}
\ \ \ $B^*(1S)$&~~~$1^{--}$&~~~$\ 5325$&~~~$\ 5325$&
~~~~~$\Lambda=75$\,MeV\\
\ \ \ $B_2^*(1P)$&~~~$2^{++}$&~~~$\ 5743$&~~~$\ 5743$&
~~~~~$m_g=416$\,MeV\\ 
\ \ \ $B_3^*(1D)$&~~~$3^{--}$&~~~$\ ~ $&~~~$\ 5946$&
~~~~~$\sigma=140$\,MeV$^2$\\ 
\ \ \ $B_4^*(1F)$&~~~$4^{++}$&~~~$\ ~ $&~~~$\ 6088$&
~~~~~$m_b=2856$\,MeV\\
\ \ \ $B_5^*(1G)$&~~~$5^{--}$&~~~$\ ~ $&~~~$\ 6205$&
~~~~~$m_d=25$\,MeV\\
\ \ \ $B^*(2S)$&~~~$1^{--}$&~~~$\ ~ $&~~~$\ 5834$&~\\
\ \ \ $B^*(2P)$&~~~$2^{++}$&~~~$\ ~ $&~~~$\ 6034$&~\\
\ \ \ $B^*(2D)$&~~~$3^{--}$&~~~$\ ~ $&~~~$\ 6175$&~\\
\noalign{\smallskip}\hline
\end{tabular}
\end{center}
\end{table}
 Note, that gluon mass $\mu_g$ is the same for glueballs~\ci{MyEPJC12}. 
 The $d$ quark effective mass decreases if heavy quark mass increases.

\section*{Conclusion}\label{Conclu}
 We have modeled mesons containing light and heavy quarks and their 
resonances in the framework of relativistic quantum mechanics. 
 We have used the modified funnel-type potential with the distant 
dependent strong coupling, $\alpha_\S(r)$.  
 Using the complex-mass analysis, we have derived the meson 
interpolating masses formula~(\ref{W2int}), in which the real and 
imaginary parts are exact expressions. 
 We have developed the free particle hypothesis: quark and antiquark 
move as free particles in the bound system. 

 We have calculated masses of light-heavy $S=1$ mesons containing 
$d$ quark and their resonances, i.\,e., $\rho^{\pm}$ and $B^{*0}$ 
states for universal string tension $\sigma=140$\,MeV$^2$, which can 
be considered as a fundamental constant in hadron physics. 
 We have found that quarks' masses are not constant values but are 
position-dependent. 
 We have studied light-heavy mesons containing $d$ quark and found  
if heavy quark mass is larger the $d$ quark mass is smaller; 
this is because distance between quarks becomes smaller. 

 This approach allows to simultaneously describe in the unified way 
the centered masses and total widths of resonances. 
 We have shown here the results only for unflavored and beauty mesons 
and resonances, however, we have obtained a~good description for 
strange and charm mesons as well. 

\bibliography{BiDaQM}

\end{document}